\documentclass[reprint,superscriptaddress,showpacs,nofootinbib,amsmath,amssymb,aps,prl]{revtex4-1}
\usepackage{graphicx}
\usepackage{dcolumn}
\usepackage{bm}
\usepackage[utf8]{inputenc}
\usepackage[frenchb]{babel}

\begin{document}

\preprint{APS/123-QED}

\title{Scale invariance and universality in a cold gas of indirect excitons}

\author{S. V. Andreev}
\email[Electronic adress: ]{Sergey.Andreev@univ-montp2.fr}
\affiliation{Laboratoire Charles Coulomb, Unité Mixte de Recherche 5221 CNRS/UM2, Université Montpellier 2, Place Eugène Bataillon, 34095 Montpellier Cedex, France}
\author{A. A. Varlamov}
\affiliation{CNR-SPIN, Tor Vergata, Viale del Politecnico 1, I-00133 Rome, Italy}
\author{A. V. Kavokin}
\affiliation{Physics and Astronomy School, University of Southampton,
Highfield, Southampton, SO171BJ, UK} \affiliation{Spin Optics Laboratory,
State University of Saint Petersburg, 1, Ulianovskaya, 198504, Russia}
\date{\today}

\begin{abstract}
We address theoretically the puzzling similarity observed in the
thermodynamic behaviour of independent clouds of cold dipolar
excitons in coupled semiconductor quantum wells. We argue that the
condensation of self-trapped exciton gas starts at the same critical temperature in all traps due to the specific scaling rule. As a consequence
of the reduced dimensionality of the system, the scaling parameters appear
to be insensitive to disorder.
\end{abstract}

\pacs{71.35.Lk}

\maketitle

\textit{Introduction}.--The ring-shaped boundary between electron rich and
hole rich regions in semiconductor quantum wells remote from the central
hot excitation spot presents the unique setting for studying of the critical
behaviour in exciton gases \cite{formation}. Indirect excitons formed on the
ring have extremely long lifetimes and high cooling rates which allow them
reaching a thermodynamic equilibrium with the cold lattice. On the other hand,
the strong repulsion of exciton dipole moments oriented perpendicularly to
the plane of the structure prevents the system from formation of biexcitons 
\cite{zimmermann} and makes possible observation of a Bose-Einstein
condensed (BEC) metastable state \cite{Keldysh}. The specifics of such BEC
can be conveniently studied by analysing the exciton photoluminescense (PL) 
\cite{Butov2007, Repulsive}.

\begin{figure}[t]
\includegraphics[width=0.8\columnwidth]{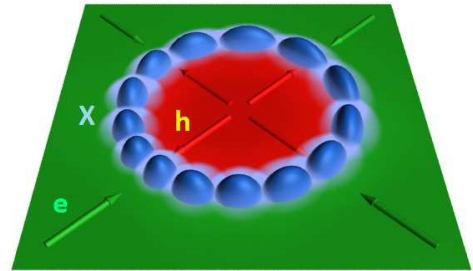}
\caption{The sketch of a pattern of exciton beads similar to one observed
by L. Butov \textit{et al.} \protect\cite{Butov2002} in CQW's structure.
Below some critical temperature the exciton ring is fragmented into an array
of independent exciton condensates (beads). The size of a bead varies weakly
along the ring due to smooth in-plane disorder. The thermal cloud of
non-condensed excitons covers all the beads.}
\label{Fig1}
\end{figure}

Intriguing phenomena have been recently observed in the PL
ring of dipolar excitons in coupled quantum wells (CQW's) \cite{Butov2002,
High2012} and, independently, in a biased single quantum well (SQW)
structure \cite{Alloing}. With lowering of temperature the exciton cloud at
the ring squeezes and fragments into an array of beads seen as bright spots in the PL spectra. Shift-interferometry measurements reveal that
each bead represents a macroscopically coherent exciton state (a
condensate). At the same time, no phase correlations between different beads
have been found \cite{High2012}. These local condensates are formed in different external
conditions, and their sizes vary along the ring [Fig. \ref{Fig1}]. Indeed,
though the electrostatic interaction between excitons results in screening
of rapid fluctuations of the in-plane potential \cite{Ivanov}, the weak
disorder varying slowly in space is always present. The
effect of disorder on the formation of patterns of exciton
condensates is a challenging problem which has not been addressed until now.

In this Letter we show that the pronounced dispersion of sizes of the beads
observed in the experiment [Fig. \ref{Fig1}] can be described accounting for
a \textit{weak} and smooth disorder potential in the system. Surprisingly, the disorder
does not affect the value of the critical temperature $T_{c}$: BEC starts in all the traps
simultaneously, at the same temperature as in the disorder free system. The
situation resembles one in multiband superconductors: in spite of the
diversity of coherence lengths and gaps of the Cooper pairs in different bands at relatively low temperatures, the system unifies close to
the phase transition, and the transition occurs at a unique critical
temperature. This important result of our model is consistent with the experemental studies \cite{Butov2002, High2012, Alloing}.

\begin{figure}[t]
\includegraphics[width=1\columnwidth]{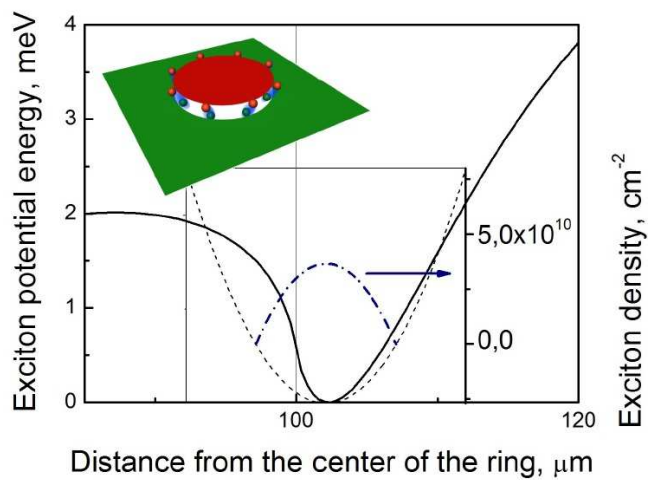}
\caption{Calculated potential profile for the radial motion of an indirect
exciton in the vicinity of the ring (solid line) and the model harmonic trap
(dashed line). Details of calculation can be found in \protect\cite{SI}. The
localization is due to the macroscopic charge separation (color inset on the
left) which induces in-plane electric field. The field tilts the exciton
dipoles and thus reduces their potential energy. At low temperatures,
excitons condense at the potential minimum located near the charge boundary.
As a consequence of strong repulsive interactions the density profile of the
exciton condensate is very smooth and merely reproduces shape of the trap
(dashed-dot line).}
\label{Fig2}
\end{figure}

In the absence of interactions and disorder, the
cloud of indirect excitons localized at the ring would condense
homogeneously at some temperature corresponding to zero chemical
potential. However, the time to reach kinetic equilibrium and build up the
long-range order in such an ideal gas would be infinitely long \cite{Pitaevskii}. The strong dipole-dipole repulsion between
excitons ensures fast thermalization of the whole cloud, but BEC occurs at a
lower critical temperature $T_{c}$ and would result in fragmentation of the ring
into a perfectly periodic array of localised condensates. In the thermodynamic limit,
the number of beads would be determined by the balance between the kinetic and
entropy terms in the free energy of the exciton system \cite{Andreev}. The exciton bead density profile along
the ring reproduces the shape of the \textit{self-trapping}
potential. The latter can be assumed to be of a harmonic type for all
beads: 
\begin{equation}
V_{j}(x,y)=\frac{1}{2}m\omega _{j}^{2}x^{2}+\frac{1}{2}m\omega _{y}^{2}y^{2},
\label{trap}
\end{equation}%
where $x$ corresponds to the azimuthal and $y$ to the radial direction, $%
j=1,2...J$ is the index of the bead and, in the absence of disorder, $\omega _{1}=\omega _{2}=...\equiv\omega
_{x}$. Localization of the clouds in the radial direction $y$ is due to the
macroscopic in-plane charge separation (see Fig. \ref{Fig2} and Ref. \cite%
{SI}). Strictly speaking, the self-trapping potential oscillates along
the ring and can be characterized by $\{\omega _{j}\}$ only in the vicinity
of its local minima. The local harmonic potential approximation \eqref{trap} is valid if the traps are deep enough and a large part of
non-condensed excitons is effectively localized in each site $j$. This is
the case in the experiments  \cite{Butov2002, High2012}. The chemical
potential of each independent cloud $\mu_{j}$ can be calculated using the normalization condition
\begin{equation}
N_{j}(\mu_{j})=\int n_{j}(x,y,\mu_{j})dxdy,  \label{norm}
\end{equation}%
where $n_{j}(x,y,\mu_{j})$ is the density profile of this cloud (see Eq. %
\eqref{density} below),
\begin{equation}
\label{N0} 
\sum\limits_{j}N_{j}(\mu_{j})=N_{0},
\end{equation}
with $N_{0}$ being the total number of excitons at the ring in a steady state, and
\begin{equation}
\label{mu}
\mu_{1}=\mu_{2}=...\equiv\mu,
\end{equation} 
so that the three equations \eqref{norm}, \eqref{N0} and \eqref{mu} determine, in fact, the \textit{unique chemical potential $\mu$ of the interacting disorder free system}.

The effect of disorder can be studied considering a perturbative correction
to the self-induced part of the localizing potential: the potential traps %
\eqref{trap} acquire different curvature (characterized by $\omega _{j}$) and contain different number of excitons $N_{j}$, while the chemical potential $\mu$ defined by \eqref{norm} remains unchanged and the condition of kinetic equilibrium \eqref{mu} is not violated. Below $T_{c}$ the latter is reached on a time scale of inverse temperature due to strong fluctuations of the relative phases between the adjacent condensates \cite{note1}. The fluctuations result in damping of coherent exciton flows \cite{note2}, which would be induced if the condition \eqref{mu} is violated. At higher temperatures the fluctuations are due to thermal activation of the phase, while close to the absolute zero they are of quantum nature and, in terms of mechanical analogy, correspond to the tunnelling of the phase between the neighboring condensates \cite{Varlamov}.

For these reasons, the condition \eqref{mu} holds for the whole range of temperatures even in the presence of disorder. Below we show, that at the experimentally achieved exciton densities this implies that the critical point $T_{c}$ is not affected by disorder and remains unique for the whole system. To obtain this non-trivial result we extend the principle of \textit{scale invariance} on a two-dimensional harmonically trapped gas and show that all the beads belong to the same \textit{universality class}.      

\textit{Scale invariance.}--Self-trapping along the ring alters dramatically
the density of exciton states making it possible to observe the true second
order phase transition in the thermodynamic limit. In this case the scale
invariance, generally taking place in the critical region \cite{Kadanoff},
was shown to be extended down to zero temperature \cite%
{Scaling, thermodynamics}. A physical reason for this specific scaling is
quenching of finite size effects (negligibility of the kinetic energy term in the
mean field equation for a condensate) \cite{Scaling}. The relevant thermodynamic functions of a trapped cloud can be expressed in terms of two parameters: the critical temperature of BEC of
non-interacting particles in a harmonic trap $T_{c,j}^{0}$ and the ratio $%
\eta_{j} =\mu_{j} (T=0)/k_{B}T_{c,j}^{0}$. In what follows we show that the scaling parameters are the same for all localized exciton clouds.

By analogy with a three dimensional problem \cite{Scaling,
thermodynamics}, the thermodynamic limit for a two dimensional harmonically
traped gas can be formally obtained by letting the total number of particles 
$N_{j}$ in a trap increase to infinity, and the oscillator frequency $\omega
_{ho,j}=(\omega _{j}\omega _{y})^{1/2}$ decrease to zero, while keeping
fixed the product $\omega _{ho,j}N_{j}^{1/2}$. The latter defines the
critical temperature of an ideal gas in a harmonic trap 
\begin{equation}
k_{B}T_{c,j}^{0}=(6/\pi ^{2})^{1/2}\hbar \omega _{ho,j}N_{j}^{1/2}.
\end{equation}

To account for contact interactions between the excitons in the trap domain,
we take advantage of the fact \cite{Pitaevskii} that as $\omega
_{ho,j}\rightarrow 0$ the density profile of the $j$th cloud $n_{j}(x,y)$ is fixed by the condition of local equilibrium 
\begin{equation}
\bar{\mu}[n_{j}(x,y),T]=\mu_{j}(T)-V_{j}(x,y).  \label{LDA}
\end{equation}%
Here $\bar{\mu}(\bar{n},T)$ is the value of a \textit{local} chemical
potential calculated for a \textit{uniform} system having the density $\bar{n%
}=n_{j}(x,y)$, while $\mu_{j}(T)$ is the chemical potential of the cloud. By inverting the condition \eqref{LDA} one can write the density $%
n_{j}(x,y)$ in the form 
\begin{equation}
n_{j}(x,y)=\bar{n}[\mu_{j}(T)-V_{j}(x,y),T],  \label{density}
\end{equation}%
where $\bar{n}(\bar{\mu},T)$ is merely the density of the uniform gas
expressed in terms of its chemical potential and temperature. At $T=0$ one
would obtain the well known Thomas-Fermi result for the condensate: 
\begin{multline}  \label{TF}
n_{j}(x,y,T=0)= \\
\frac{1}{V_{0}}[\mu_{j}(T=0)-V_{j}(x,y)]\theta[\mu_{j}(T=0)-V_{j}(x,y)],
\end{multline}
where $\theta(x)$ is the Heaviside step function. In practice, tracing the
local value of the chemical potential $\bar{\mu}(x,y)$ allows one to
reproduce the density profile of a cloud and vice versa \cite{Nature}.

Using the normalization condition \eqref{norm} and Eq. \eqref{TF} one finds
the chemical potential at $T=0$ in the form 
\begin{equation}
\mu_{j}(T=0)=\sqrt{\frac{mV_{0}}{\pi\hbar ^{2}}}\hbar\omega_{ho,j}N_{j}^{1/2},
\label{muTF}
\end{equation}%
and the ratio 
\begin{equation}
\eta_{j} \equiv\frac{\mu_{j}(T=0)}{k_{B}T_{c,j}^{0}}=\sqrt{\frac{\pi }{6}\frac{mV_{0}}{\hbar ^{2}}}.  \label{eta}
\end{equation}
\textit{Crucially, in contrast to the case of a
three dimensional gas \cite{Scaling} the quantities $\eta_{j}$ in \eqref{eta} are
independent on the oscillator frequencies $\omega _{j}$ and $\omega _{y}$
characterizing the trap and on the number of particles in a cloud $N_{j}$.} Providing that the condition \eqref{mu} is satisfied, this implies that the critical temperature of an ideal gas $T_{c,j}^{0}$ is also the same for all traps, so that one can write 
\begin{subequations}
\label{parameters}
\begin{align}
\eta_{1}&=\eta_{2}=...\equiv\eta,\\
T_{c,1}^{0}&=T_{c,2}^{0}=...\equiv T_{c}^{0}.
\end{align}
\end{subequations}

In order to show that the quantities $\eta$ and $T_{c}^{0}$ are the scaling parameters, we follow Ref. \cite{Pitaevskii} and introducing a new variable $\xi \equiv V_{j}(x,y)$ rewrite the identity %
\eqref{norm} in the form 
\begin{equation}
\label{integral}
2(k_{B}T_{c}^{0})^{-2}\int \frac{6}{\pi }\frac{\hbar ^{2}}{m}\bar{n}(\mu
-\xi ,T)d\xi =1,
\end{equation}
where we have used Eq. \eqref{density} with $\mu_{j}$ replaced by $\mu$ according to \eqref{mu}. Inversion of the equation \eqref{integral} yields the general
dependence $\mu =\mu (T,T_{c}^{0},\eta )$ for the chemical potential of the
trapped cloud. Due to the dimensionality arguments this expression can be
recast in the form 
\begin{equation}
\label{scaling}
\mu =k_{B}T_{c}^{0}f(t,\eta ),
\end{equation}
where $t\equiv T/T_{c}^{0}$ is the reduced temperature, $f$ is a generic
function which satisfies $f(0,\eta )=\eta$.

Equation \eqref{scaling} exhibits the anticipated scaling in terms of $\eta$
and $T_{c}^{0}$. By analogy, one can show the scaling of all other
thermodynamic functions. Having in mind the result \eqref{parameters} one can conclude that all the beads belong to the same universality class defined by $\eta$
and $T_{c}^{0}$. In particular, the critical point is unique for the whole system even in the presence of disorder. To illustrate this important result, let us estimate $T_{c}(\eta,T_{c}^{0})$ for a small $\eta$, where the simplest Hartree-Fock scheme can be applied \cite{thermodynamics}. In this approximation,
\begin{equation}
\label{f}
f(t,\eta)=\eta(1-t^{2})^{1/2}
\end{equation}
and $T_{c}$ can be found solving the transcendental equation $\mu(T_{c},T_{c}^{0},\eta)=\epsilon[\mu(T_{c},T_{c}^{0},\eta),T_{c}]$ with $\epsilon$ being the lowest eigenvalue of the single particle Hamiltonian $H_{\mathrm{sp},j}=V_{j}(x,y)+2V_{0}n_{j}(x,y)$. Using Eqs. \eqref{density}, \eqref{scaling}, and \eqref{f} one finds \cite{SI}
\begin{equation}
\label{tc}
T_{c}=T_{c}^{0}\left(1+\frac{x^2(\eta)}{\eta^{2}}\right)^{-1/2},
\end{equation}
where $x(\eta)$ is a root of $\pi^{2}x=6\eta^{2}\mathrm{Li}(e^{-x})$, $\mathrm{Li}(x)$ is the Eulerian logarithmic integral \cite{Handbook}.

The scaling arguments given above are based on the local density approximation (LDA) given by Eq. \eqref{density} or, equivalently, \eqref{LDA}. Experimental \cite{Nature} and \textit{ab initio} \cite%
{abinitio} studies show that LDA for 2D gases is already valid to a good accuracy for $\sim 10^{4}
$ particles. This corresponds to the experimentally achieved exciton
densities in a bead \cite{Butov2002}. However, since an exciton gas is quite different from usual atomic gases, it is worth to discuss the applicability of LDA for the beads in details. 

\textit{The validity of the local density approximation for the beads.}--To verify the validity of the Thomas-Fermi
approximation for exciton clouds, we notice that the local chemical
potential $\bar{\mu}(x,y)$ given by Eq. \eqref{LDA} can be inferred from the PL energy profiles along the ring measured in \cite%
{Repulsive} and shown there in Fig. 2. Indeed, the chemical potential $%
\bar{\mu}(x,y)$ contributes to the energy of a photon emitted by an
exciton during recombination. Neglecting the thermal component of the
exciton gas, the average PL energy measured in \cite{Repulsive} for one bead
can be written as 
\begin{equation}
E_{\mathrm{PL}}(x)-const=\frac{\int \bar{\mu}(x,y)n(x,y)dy}{\int
n(x,y)dy}=\frac{4}{5}\bar{\mu}(x,y=0)  \label{Epl}
\end{equation}%
with $n(x,y)$ given by \eqref{TF} where we have omitted the
index $j$ for simplicity (this change in the notation will be kept until the end of this
section). We choose a bead in the middle of
Fig. 2 of Ref. \cite{Repulsive}, which has the most regular shape compared
to its neighbours. When expressed in reduced units and multiplied by 5/4
to account for the averaging along $y$ axis [Eq. \eqref{Epl}], the energy
profile of this bead reads $\bar{\mu}/k_{B}T_{c}^{0}=\eta -\tilde{x}^{2}$ [{Fig. \ref{Fig3}], where 
$\tilde{x}=\eta ^{1/2}x/R_{x}$ and 
\begin{equation}
R_{x}=[2\mu (T=0)/m\omega_{x}^{2}]^{1/2}  \label{TFradius}
\end{equation}%
is the Thomas-Fermi radius. Here we have substituted $\mu (T=0)=\eta
k_{B}T_{c}^{0}$ into the right-hand side of Eq. \eqref{LDA}. Note, that we do not adjust the scaling parameters: we find $\eta
=1.6$ using Eq. \eqref{eta}, where we substitute $V_{0}=1.7$ $\mu eV\times
\mu m^{2}$ calculated using the plate capacitor formula with the correction
factor \cite{zimmermann}. In what concerns the parameter $T_{c}^{0}$, it can
be estimated from the
experimental temperature at which the fragmentation and the
build up of the extended coherence occur $T_{c}^{0}=4.5$ K \cite{Repulsive}. 
\begin{figure}[t]
\includegraphics[width=1\columnwidth]{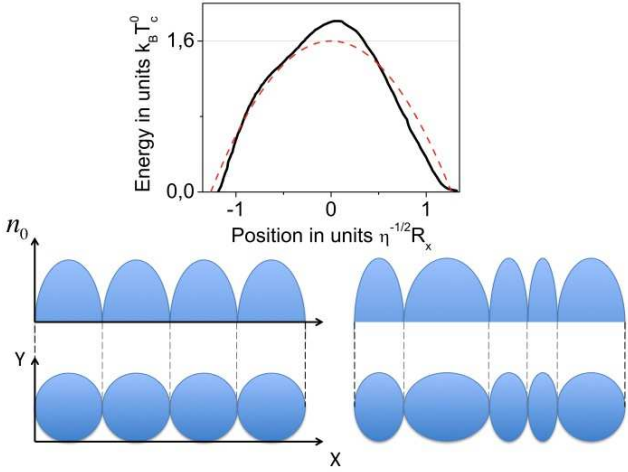}
\caption{\textit{(top)} The $T=0$ Thomas-Fermi result $\bar{\protect\mu}%
/k_{B}T_{c}^{0}=\protect\eta -\tilde{x}^{2}$ for the variation
of exciton energy along the ring ($\tilde{x}$ axis) at $\tilde{y}=0$ (the dashed
red line). The exciton resonance position measured in \cite{Repulsive} from PL spectra is shown by the solid line. The scaling parameters are $\protect\eta =1.6$
and $T_{c}^{0}=4.5$ K. The Thomas-Fermi radius of the bead is measured to be 
$R_{x}=20$ $\protect\mu $m. \textit{(Bottom)} The topological transformation
of the condensate density given by \eqref{transform} conserving the total
number of particles $N_{0}=\protect\int n_{0}dxdy$ and the Thomas-Fermi
energy $E_\mathrm{TF}=V_{0}\protect\int n_{0}^{2}dxdy$.}
\label{Fig3}
\end{figure}
}

Furthermore, using the Thomas-Fermi approximation \eqref{TFradius} for
available values of the parameters $\eta $ and $T_{c}^{0}$ one can estimate
the oscillator frequency $\omega _{y}$ of the radial localizing potential
which would correspond to the experimentally observed ring width $2R_{y}\sim
10$ $\mu$m. Remarkably, this potential can be obtained from the first
principles, see \cite{SI}. The result of this calculation is shown in Fig. %
\ref{Fig2} by solid line. The input parameters for the calculation procedure
correspond to those typical of the experiment. Dashed line shows the model
harmonic potential. We plot also the ground state density profile for $\eta
=1.6$ and $T_{c}^{0}=4.5$ K (we have assumed $\omega _{x}=\omega _{y}$ for
simplicity). As one could expect, the semiclassical condition $\hbar \omega
_{y}\ll k_{B}T_{c}^{0}$ is well satisfied.

\textit{The energy scale of the disorder}--Finally, let us estimate the energy scale of the disorder potential which can induce the significant dispersion of bead sizes observed in practice [Fig. \ref{Fig1}]. We do not wish to complicate the issue by taking into account the thermal
component of the gas and, therefore, consider the fragmented exciton
condensate at $T=0$. As we have already explained, in the scaling regime
this restriction does not imply any loss of generality.

It is reasonable to assume that at $T=0$ the adjacent condensates touch each other as it is shown schematically in Fig. \ref{Fig3} (in order to minimize the interaction energy $E_\mathrm{TF}$). In the Thomas-Fermi limit this means that the oscillator frequencies $\{\omega_{j}\}$ satisfy the "continuity" condition
\begin{equation}
\label{condition}
\sum_{j}[2\mu(T=0)/m\omega_{j}^{2}]^{1/2}=\pi R,
\end{equation}
where $R$ is the ring radius. The smooth disorder can fragment the condensate and make varying the bead sizes along the ring, while conserving the parabolic shape of the bead density profiles $n_{j}(x,y,T=0)$. Interestingly, such topological transformation of the exciton density can be formally achieved by the replacement 
\begin{equation}
\{\omega_{j}\}\rightarrow \{\omega_{k}\}^{\ast },  \label{transform}
\end{equation}%
where $\{\omega_{k}\}^{\ast }$ is a new set of oscillator frequencies, $k=1,2...K$, satisfying the condition \eqref{condition} with $k$ instead of $j$ and $K\neq J$ in general. One can
check \cite{SI} that the transformation \eqref{transform} conserves the total number
of particles $N_{0}$ [Eq.\eqref{N0}]. 

This way, one can achieve the pronounced dispersion of the bead sizes observed experimentally maintaining the chemical potential $\mu(T=0)$ corresponding to \textit{the disorder free system}. This suggests that the variation of the disorder potential $\delta$ on the scale of the bead size is much less than $\mu(T=0)$. Indeed, not only the sum $\sum_{j} \int n_{j}(x,y,T=0)dxdy=\mathrm{inv}$
but also
\begin{equation}
\label{sum}
\sum\limits_{j} \int n_{j}^{2}(x,y,T=0)dxdy=\mathrm{inv}
\end{equation}
under the topological transformation defined by \eqref{transform}. Equation \eqref{sum} defines the energy accumulated in the clouds due to
the repulsive interaction (the Thomas-Fermi energy) $E_\mathrm{TF}=V_{0}\sum_{j} \int n_{j}^{2}(x,y,T=0)dxdy$. To estimate the lowest bound for $\delta$ one
should go beyond the scaling limit. It is shown in \cite{Andreev} that the kinetic energy correction to the Thomas-Fermi approximation can be estimated as $k_{B}T_{c}^{0}/\eta$ (per one bead). Therefore, it is sufficient to introduce a weak disorder which varies smoothly by 
\begin{equation}
k_{B}T_{c}^{0}/\bar{N_{j}}\eta<\delta\ll\mu(T=0)\equiv\eta k_{B}T_{c}^{0}
\end{equation}
on the scale of the bead size so that one could observe its effect
on the fragmented exciton condensate ($\bar{N_{j}}$ is the average number of particles in a bead). The high sensitivity of condensate sizes to the disorder reflects the fact that the trapping potential along the ring is essentially self induced.  

\textit{Conclusions.}--We have shown that the fragmented
exciton ring represents an array of trapped Bose-Einstein condensates close
to the thermodynamical limit. The relevant thermodynamic functions of
exciton clouds exhibit scaling in terms of the parameters $\eta$ and $T_{c}^{0}$. With lowering temperature the lakes of condensed excitons grow maintaining the same chemical potential. The dispersion of their sizes reveals weak and
smooth structural disorder, which is hidden from an observer above $%
T_{c}$. As a consequence of the reduced dimensionality, such disorder does not alter the scaling parameters. This
explains the experimentally observed universality in the thermodynamic
behaviour of statistically independent exciton condensates.

The work has been supported by EU ITN project "CLERMONT 4". S. V. is
grateful to K. V. Kavokin for fruitful discussion on the existence of the
in-plane radial trap for excitons. A.V. acknowledges the support of the EU
IRSES project SIMTECH No. 246937, A.K. acknowledges support from the Russian Ministry
of Education and Science, grant N11.G34.31.0067.


\begin{thebibliography}{1}
\bibitem{note1} On a time scale of the exciton radiative lifetime the constant chemical potential along the ring $\mu$ is established  due to the dynamical equilibrium between localized excitons and delocalized electron-hole plasma.
\bibitem{note2} Such damping mechanism is well known in the physics of superconductors, where the fluctuations of the phase in Josephson junctions result in energy dissipation and decay of supercurrents \cite{Varlamov}.
\end{thebibliography}
\end{document}


\title{Supplemental information\\"Scale invariance and universality in a cold gas of indirect excitons"}

\author{S. V. Andreev}
\email[Electronic adress: ]{Sergey.Andreev@univ-montp2.fr}
\affiliation{Laboratoire Charles Coulomb, Unité Mixte de Recherche 5221 CNRS/UM2, Université Montpellier 2, Place Eugène Bataillon, 34095 Montpellier Cedex, France}
\author{A. A. Varlamov}
\affiliation{CNR-SPIN, Tor Vergata, Viale del Politecnico 1, I-00133 Rome, Italy}
\author{A. V. Kavokin}
\affiliation{Physics and Astronomy School, University of Southampton,
Highfield, Southampton, SO171BJ, UK} \affiliation{Spin Optics Laboratory,
State University of Saint Petersburg, 1, Ulianovskaya, 198504, Russia}

\maketitle
\section{A 2D ideal gas in a harmonic trap}
Consider two-dimensional non-interacting (ideal) gas in a harmonic trap $V_{ext}(x,y)=\frac{1}{2}m\omega_{x}^{2}x^{2}+\frac{1}{2}m\omega_{y}^{2}y^{2}$. Let it be close to the thermodynamic limit. Then one can use the semiclassical relation for density of states (DOS)
\begin{equation}
g(\epsilon)=\int\frac{d^2 r d^2 p}{(2\pi\hbar)^2}\delta(\epsilon-p^{2}/2m-V_{ext}(x,y))
\end{equation}
The integration yields $g(\epsilon)=\epsilon/(\hbar\omega_{ho})^2$, where $\omega_{ho}=(\omega_{x}\omega_{y})^{1/2}$.

The number of particles in the thermal cloud below $T_{c}^{0}$ (the chemical potential $\mu=0$)
\begin{equation}
N_{T}=\int d\epsilon\frac{g(\epsilon)}{e^{\beta\epsilon}-1}
\end{equation}
The condition $N_{T}=N$ yields the following equation for the critical temperature
\begin{equation}
k_{B}T_{c}^{0}=(6/\pi^{2})^{1/2}\hbar\omega_{ho}N^{1/2}
\end{equation}
\section{2D Bose-Einstein condensate with repulsive interactions at $T=0$}
Bose-Einstein condensate of an interacting gas can exist in a 2D trap. As in the previous section, consider a system of bosons close to the TD limit. This allows one to neglect the kinetic energy term in GP equation and to calculate the relevant functions of the system in the Thomas-Fermi approximation.

The density profile
\begin{equation}
n(x,y)=\frac{1}{V_{0}}(\mu-V_{ext})
\label{rho}
\end{equation}
if $\mu>V_{ext}$ and is zero elsewhere. The chemical potential
\begin{equation}
\mu(T=0)=\sqrt{\frac{mV_{0}\omega_{x}\omega_{y}}{\pi}}N^{1/2},
\label{muTF}
\end{equation}

Useful parameter is the ratio $\eta=\mu(T=0)/k_{B}T_{c}^{0}$. One obtains
\begin{equation}
\eta =\sqrt{\frac{\pi }{6}\frac{mV_{0}}{\hbar ^{2}}}.  
\label{eta}
\end{equation}

In the thermodynamic limit the size of the condensate is limited by 
a boundary set by the condition $\mu=V_{ext}(x,y)$. This is valid also for $T>0$. Noteworthy, the peak condensate density $n(x=0,y=0)$ is fixed by $\mu$.

\section{Topological transformation of the fragmented condensate conserving the total number of particles and the chemical potential}

In this section, we show that the transformation of the fragmented condensate density profile $n_{0}\rightarrow n_{0}^{*}$ defined by Eq. (19) in the main text [we have introduced $n_{0}(x,y)=\sum_j n_{j}(x,y)$] conserves simultaneously the integrals
\begin{equation}
\begin{split}
J_{1}& =\int n_{0} dxdy\\
J_{2}& =\int n_{0}^{2} dxdy
\end{split}
\end{equation}

For that purpose let us calculate explicitly the volume $V$ of the following object in $(x,y,z)$ space:
\begin{equation}
\label{figure}
\frac{z}{h}=\left(\frac{x}{R_{x}}\right)^{2\alpha}+\left(\frac{y}{R_{y}}\right)^{2\alpha}+\beta\left(\frac{xy}{R_{x}R_{y}}\right)^{\alpha},
\end{equation}
where $(\alpha=1,\beta=0)$ or $(\alpha=2,\beta=2)$.
Introducing cylindrical coordinates $x=r\cos\theta$ and $y=r\sin\theta$ we write the area $S(z)$ of cross-section at the height $z$ in the form
\begin{equation}
S(z)=\int\limits_{0}\limits^{2\pi}\int\limits_{0}\limits^{r(z)}rdrd\theta=\left(\frac{z}{h}\right)^{1/\alpha}R_{x}R_{y}\int\limits_{-\infty}\limits^{+\infty}(1+x^{2\alpha}+\beta x^{\alpha})^{-1/\alpha}dx
\end{equation}
so that the volume reads as
\begin{equation}
V=\int\limits_{0}\limits^{h}S(z)dz=\frac{R_{x}R_{y}h}{1/\alpha+1}\int\limits_{-\infty}\limits^{+\infty}(1+x^{2\alpha}+\beta x^{\alpha})^{-1/\alpha}dx.
\end{equation}
One can see that the resulting expression is linear in $R_x$. Consequently, the volume of our object is equal to the sum of volumes $V_j$ of J objects (j=1,2,..., J) described by Eq. (8) with the fixed $h$ and $R_y$ and different $R_{x,j}$ if the sum of $R_{x,j}$ equals $R_x$:
\begin{equation}
\label{partition}
V=V_{1}+V_{2}+...+V_{J}
\end{equation}
The possibility of the decomposition \eqref{partition} proves the statement at the beginning of this section. Indeed, by definition the topological transformation of the exciton density defined by Eq. (19) satisfies the condition (18) (see the main text), which is nothing but the condition $\sum_j R_{x,j}=\pi R$ [compare (18) with Eq. (17)].

\section{The critical temperature of a weakly interacting 2D gas in a harmonic trap}

Let us derive Eq. (15) for the critical temperature of a weakly interacting gas in a 2D harmonic trap. Due to the presence of the trap, the thermal density is much smaller than the condensate density even if the fraction of uncondensed excitons is comparable with $N_{j}$. As a consequence, the Thomas-Fermi result \eqref{muTF} can be straightforwardly extended to finite temperatures
\begin{equation}
\label{muT}
\mu(T)=\mu(T=0)\left(\frac{N_{c}}{N}\right)^{1/2}=\eta k_{B}T_{c}^{0}(1-t^{2})^{1/2},
\end{equation}
where we have used the ideal gas result $N_{c}=N(1-t^{2})$ for the number of excitons in the condensate.   
Equation \eqref{muT} is valid for a weakly interacting gas, where the condition $\mu(T)<k_{B}T_{c}^{0}$ is satisfied.

Bose-Einstein condensation starts at the temperature $T_{c}$ for which the chemical potential \eqref{muT} reaches the lowest eigenvalue $\epsilon$ of the single-particle Hamiltonian, which in the simplest Hartree-Fock approximation can be recast as
\begin{equation}
\label{H}
\hat{H}_\mathrm{sp}=-\frac{\hbar^{2}\mathbf{\nabla}^{2}}{2m}+V_{ext}(x,y)+2V_{0}n(x,y).
\end{equation}
In the scaling regime one can safely neglect the kinetic energy term for the ground state of \eqref{H} and, using the semiclassical expression
\begin{equation}
\label{thermal}
n(x,y)=\int\frac{d\mathbf{p}}{(2\pi\hbar)^{2}}\left[\exp\left(\frac{H_\mathrm{sp}(p^{2},x,y)-\mu}{k_{B}T}\right)-1\right]^{-1} 
\end{equation}
rewrite the condition $\mu(T_{c})=\epsilon(T_{c})$ in the form (15) (see the main text), which holds for small $\eta$.

\section{Built-in electrostatic trap for indirect excitons}

Since the first experimental realization of a cold exciton gas in CQW several trapping configurations have been proposed and realized \cite{nanoletters, Gippius, Gorbunov, Rapaport1}. These proposals rely on the electrostatic interaction of the exciton dipole moment with collinear electric field $E_{z}$ ($z$ is the structure growth direction). The interaction energy is given by $\varepsilon(x,y)=-edE_{z}(x,y)$, showing that the excitons will seek for the regions in the $(x,y)$ plane where the electric field is stronger. The opportunity to realize the desired potential landscape as well as the ability to control it on a time scale shorter than the exciton lifetime opens large perspectives in manipulation of excitonic condensates.

Until now it has been assumed, that the excitons created on the ring diffuse out as nearly free particles (if one neglects the exciton-exciton interaction) \cite{formation, Haque}. Here we show that at the carrier densities achieved in this kind of experiment the radial motion of the dipolar excitons is confined by a spontaneously induced in-plane potential well, i. e. the excitons are localized at the ring. The origin of this well can be understood from the following.

As we have already mentioned, the exciton ring appears on the boundary between the internal hole-rich region and the external electron-rich region. Thus one can expect the existence of in-plane electric field $\textbf{E}_{r}(r,\theta)$ in the vicinity of the ring oriented everywhere outwards the center, along the radii. In the ground state an exciton does not interact with this field, since the relative two-dimensional motion of the electron-hole pair is characterized by a symmetric wavefunction \cite{Ivchenko}. However, the interaction is already possible in the first order of the perturbation theory due to admixing of the $2p$-exciton state. Classically, the electric field "stretches" the exciton, tilting its dipole moment. As in the case of the artificially created potential profiles discussed above, the interaction of an in-plane component of the exciton dipole moment with the built-in electric field at the boundary results in appearance of an electrostatic trap for an exciton in the radial direction, which localizes the exciton gas on the ring.

Following the general rules of quantum mechanics one can obtain
\begin{equation}
\label{epsilon}
\varepsilon(r)=-\frac{\vert ea_{2D}E_{r}(r)\vert^{2}}{\varepsilon_{2p}-\varepsilon_{1s}}
\end{equation}
for the corresponding interaction energy, where $a_{2D}$ is the 2D effective exciton Bohr radius. The denominator can be estimated as the energy difference between the $1s$ ground state and the $2s$ excited state of the indirect exciton. These have been calculated in Ref. \onlinecite{Muljarov} and can be easily identified in Fig. 4 therein. Our main goal is to calculate the radial electric field $E_{r}$ as a function of the distance $r$ from the center of the ring.

\begin{figure}[tbp]
\includegraphics[width=0.5\columnwidth]{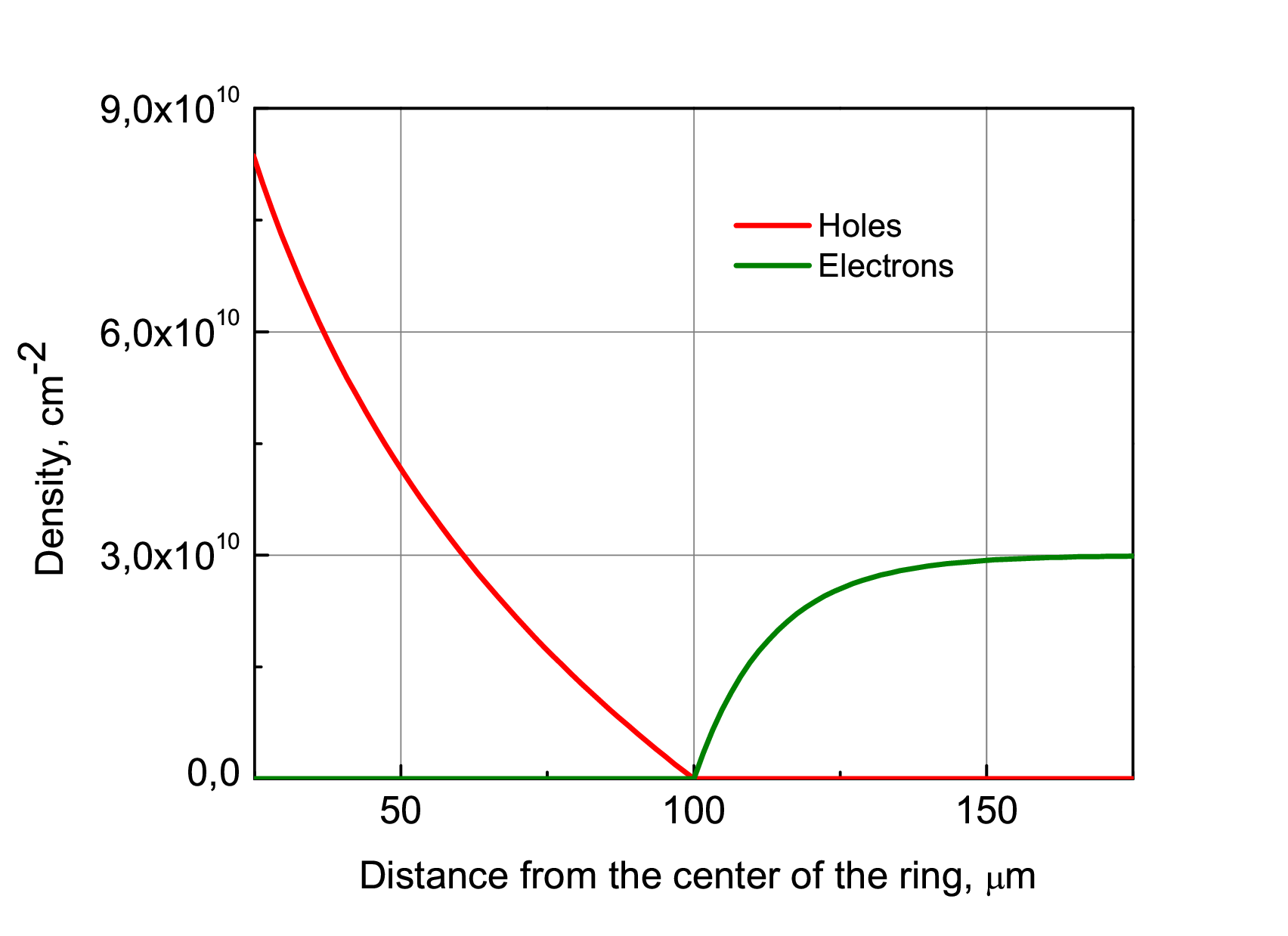}
\caption{The carrier densitity distribution in the radial direction according to the analytic expressions \eqref{densities}. The parameters $p$ and $n$ were adjusted to produce the trap shown in Fig. 2 in the main text, which is close to the model parabolic trap discussed therein. }
\label{Fig4}
\end{figure}

For that purpose one needs to know the electron and the hole density distributions, $n_{e}$ and $n_{h}$.  These can be obtained from the reaction-diffusion model supplemented with the drift term due to the induced electric field. Here we start with the simplest model proposed in Ref. \onlinecite{Haque}, neglecting the interaction with the electric field and allowing the analytic expressions for $n_{e}$ and $n_{h}$:
\begin{subequations}
\label{densities}
\begin{align}
n_{h}(x)& =p\ln(x^{-1})&
0<x\leqslant 1\\
n_{e}(x)& =n\left(1-x^{-1/2}e^{-\zeta(x-1)}\right)&
x>1
\end{align}
\end{subequations}
where we have introduced $x\equiv r/R$, $\zeta=R/l$, $R$ is the ring radius and $l$ is the electron depletion length. We assume $R\gg l$ since in the same limit the ring radius is predicted to change linearly with the laser excitation power \cite{Haque}, that has been indeed observed experimentally, see Ref. \cite{Andreev}. The carrier density profiles \eqref{densities} are shown in Fig. \ref{Fig4}.

The further simplification is to neglect the curvature of the interface. The numerical study shows that this is indeed correct at least for large enough ring radii (compared with the width of the ring) and when one is interested in the field distribution in the vicinity of the interface [the latter corresponds to $x=1$ in \eqref{densities}].

In this simplified geometry the total electric field at the point $y\equiv r/R$ can be represented as the superposition of electric fields from a large number of elementary filaments parallel to the interface:
\begin{equation}
\label{Er}
E_{r}(y)=\frac{e}{2\pi\epsilon\epsilon_{0}}\left(\int\limits^{1}\frac{n_{h}(x)dx}{y-x}-\int\limits_{1}\frac{n_{e}(x)dx}{y-x}\right)
\end{equation}
The calculation of integrals in Eq. \eqref{Er} is not trivial. The integrals have logarithmic divergencies at the point we are interested in and should be regularized in the sense of the principal value. The resulting set of equations, suitable for numerical calculations, has the form
\begin{widetext}
\begin{subequations}
\label{field}
\begin{align}
E_{r}(y)&=\frac{e}{2\pi\epsilon\epsilon_{0}}\left(\int\limits_{0}^{\lvert y-1\rvert}\frac{\pm n_{e,h}(y+x)\mp n_{e,h}(y-x)}{x}dx+\int\limits_{\lvert y-1\rvert}^{x_{0}}\frac{n_{h}(y-x)+n_{e}(y+x)}{x}dx\right)&
\lvert y-1\rvert\leqslant x_{0}\\
E_{r}(y)&=\frac{e}{2\pi\epsilon\epsilon_{0}}\int\limits_{0}^{x_{0}}\frac{\pm n_{e,h}(x+y)\mp n_{e,h}(x-y)}{x}dx&
\lvert y-1\rvert>x_{0}
\end{align}
\end{subequations}
\end{widetext}
where the sign $+$/$-$ as well as the subscript $e$/$h$ states for $y>1$ or $y\leqslant 1$ respectively.

Finally, one should bear in mind, that the CQW's are surrounded by conducting contact layers \cite{Butov2007}. The contacts will contain electrostatic images of the electron and the hole lakes, located on the distances $2kh$ from the QW's, where $h$ is the barrier width (we neglect the space between the QW's) and $k=1,2...$ is an integer. The contribution from the images can be taken into account by replacing
\begin{equation}
n_{e,h}\longrightarrow n_{e,h}\left(1+2x\sum\limits_{k=1}^{\infty}\frac{(-1)^{k}}{\sqrt{x^{2}+k^{2}\xi^{2}}}\right)
\end{equation}
in the integrals \eqref{field}, where $\xi\equiv 2h/R$.

The exciton potential energy profile calculated using Eqs. \eqref{epsilon} and \eqref{field} is presented in Fig. 2 in the main text. We used the following set of parameters: $p=6\times10^{10}$ $cm^{-2}$, $n=3\times10^{10}$ $cm^{-2}$, $\epsilon=12.9$, $a_{2D}=20$ nm, $h=200$ nm, $R=100$ $\mu$m, $l=14$ $\mu$m and $(\varepsilon_{2p}-\varepsilon_{1s})=2$ meV. The upper bound $x_{0}$ in the integrals \eqref{field} was increased until the shape, the effective width and the height of the trap saturated. This criterion is already achieved at $x_{0}\sim 0.2$, that justifies the simplified geometry we have used. Further increase of $x_{0}$ leads to the shift of the whole picture towards higher energies. Since we are only interested in the motion of the excitons in the vicinity of the ring, we conveniently set the origin of the energy axis at the bottom of the trap.

We point out that even in this simplified model one has a large set of parameters. As expected, the depth of the trap is strongly sensitive to the electron and hole densities. The latter can be roughly estimated from the laser excitation density (see Ref. \cite{Snoke}). The background electron density can in principle be determined by measuring the electron current through the sample in the absence of photoexcitation \cite{Yang}, provided that the electron escape time from the QW due to tunneling is known in advance \cite{Haque}. To the best of our knowledge, neither the electron current nor the tunneling time haven't been measured yet. Varying the electron and hole densities near some realistic values, we could obtain trapping configuration which correspods to the experimentally measured size of a condensate and the parameter $T_{c}^{0}$.